# 3D scaffold with effective multidrug sequential release against bacteria biofilm

Rafaela García-Alvarez[1], Isabel Izquierdo-Barba[1,2]*, María Vallet-Regí[1,2,]*

1. Dpto. Química Inorgánica y Bioinorgánica. Universidad Complutense de Madrid. Instituto de Investigación Sanitaria Hospital 12 de Octubre i+12. Plaza Ramón y Cajal s/n, 28040 Madrid, Spain.
2. CIBER de Bioingeniería, Biomateriales y Nanomedicina, CIBER-BBN, Madrid, Spain

* Corresponding authors. E-mail address: ibarba@ucm.es; vallet@ucm.es;
Phone: +34913941861  Fax: +34 394 17 86

**Abstract**

Bone infection is a feared complication following surgery or trauma that remains as an extremely difficult disease to deal with. So far, the outcome of therapy could be improved with the design of 3D implants, which combine the merits of osseous regeneration and local multidrug therapy so as to avoid bacterial growth, drug resistance and the feared side effects. Herein, hierarchical 3D multidrug scaffolds based on *nanocomposite* bioceramic and polyvinyl alcohol (PVA) prepared by rapid prototyping with an external coating of gelatin-glutaraldehyde (Gel-Glu) have been fabricated. These 3D scaffolds contain three antimicrobial agents (rifampin, levofloxacin and vancomycin), which have been localized in different compartments of the scaffold to obtain different release kinetics and more effective combined therapy. Levofloxacin was loaded into the mesopores of nanocomposite bioceramic part, vancomycin was localized into PVA biopolymer part and rifampin was loaded in the external coating of Gel-Glu. The obtained results show an early and fast release of rifampin followed by sustained and prolonged release of vancomycin and levofloxacin, respectively, which are mainly governed by the progressive in vitro degradability rate of these scaffolds. This combined therapy is able to destroy Gram-positive and Gram-negative bacteria biofilms as well as inhibit the bacteria growth; in addition, these multifunctional scaffolds exhibit excellent bioactivity as well as good biocompatibility with complete cell colonization of preosteoblast in the entire surface, ensuring good bone regeneration. These findings suggest that these hierarchical 3D multidrug scaffolds are promising candidates as platforms for local bone infection therapy.

**Keywords:** Multidrug 3D scaffold, combined therapy, biofilm, Gram-positive bacteria, Gran-negative bacteria and sequential antimicrobial delivery.

# 1. Introduction

Bone infection is a potentially devastating complication with important clinical and socio-economic implications [1-3]. It is described as an inflammatory process that leads to bone destruction (osteolysis) usually caused by an underlying microbial infection, mainly by *Staphylococcus aureus* bacteria [4,5]. Conventional treatments, involving systemic antibiotic administration [6], surgery and implant removal [7,8], have important limitations and significant repercussions for the quality life of the patients such as, high side effects [9], prolonged hospital stays, additional surgical interventions [10], and even, high morbidity rate [11]. The main reason for the failure of these conventional treatments is the ability of bacteria to develop a biofilm [1-4]. *Biofilms* are described as communities of microorganisms that grow attached to a surface or interphase and embedded in a self-produced extracellular matrix [12]. *Biofilm* development is one of the most common processes that bacteria accomplish in a cooperative manner. Inside the *biofilm*, bacteria grow protected from environmental stresses and resist antibiotics, disinfectants, phagocytosis and other components of the innate and adaptive immune and inflammatory defense system of the host giving as a consequence failure of the antibiotic treatment [13].

Currently, the nanotechnology field has emerged as a powerful tool to combat the infection process [14-16]. In this sense the development of novel multifunctional 3D scaffolds based on nanostructured materials to not only clear the infection but to also contribute to subsequent bone regeneration would be a very good alternative to conventional therapies [17-23]. The local antimicrobial administration will minimize side effects and risk of overdose, as well as to improve the bioavailability of the drug with the appropriate therapeutic concentration effectively reaching the target site [24]. Moreover, the possibility to achieve a combined therapy with different antimicrobial agents would be needed for a more efficient treatment of bone infection [25]. In this sense, an initial fast release of an antibiofilm drug to be able to destroy the *biofilm capsule* and subsequently more sustained and prolonged release of the different antibiotics would be desired [26].

Recently, a nanocomposite bioceramic (MGHA), formed by particles of nanocrystalline apatite embedded into amorphous mesoporous bioactive glass in the $SiO_2$–$P_2O_5$–$CaO$ system, has been reported. Due to the synergy of the features of its two components, including (i) ordered mesoporous arrangement with pores of 8 nm, (ii) high surface area and pore volume, (iii) high bioactivity, (iv) presence of nanocrystalline apatite particles homogeneously distributed, and (v) improved in vitro biocompatibility, this nanocomposite material is an excellent candidate for bone tissue engineering and local drug delivery [27-30]. Concerning the fabrication processes to obtain scaffolds, 3D plotting techniques (also called direct writing or printing) have been widely developed to prepare porous scaffolds in recent years [31]. In this sense, recently, hierarchical meso-macro 3D porous scaffolds have been fabricated by a combination of a single-step sol–gel route in the presence of a surfactant as the mesostructure directing agent and a biomacromolecular polymer (methylcellulose) as the macrostructure template followed by rapid prototyping technique with a high potential in bone tissue engineering [32,33]. However, this methodology for preparing scaffolds containing different antimicrobial agent is inconvenient, because of the need for methylcellulose and the additional sintering procedure.

Currently, the fabrication of *composite scaffolds* based on bioceramic-polymer mixture has allowed improvement of their properties due to the synergy of the features of their components being widely used for different applications [34,35]. Therefore, the incorporation of biocompatible polymer have enhanced the mechanical properties of scaffolds by using room temperature in the process, which offers the possibility to incorporate drugs for the subsequently controlled release [36]. In this case, polyvinyl alcohol (PVA) is selected because it is generally biocompatible, degradable and water soluble so toxic solvents do not need to be used in the preparation. In addition, PVA can be cross-linked to improve its crystallinity and to control its dissolution by a simple heat treatment at low temperature [37]. A previous study has shown that mixing mesoporous glasses powders with an aqueous PVA solution to form an injectable paste is very efficient to fabricate 3D scaffolds. In addition, several studies have shown that gelatin-glutaraldehyde (Gel-Glu) coatings onto the 3D scaffolds could improve both, its mechanical properties as well as the early and fast release of a drug depending on the cross-linking degree of the gelatin [38-40].

The present study is focused in finding an adequate therapeutic solution for the treatment of bone infection by the design of hierarchical 3D multidrug scaffolds based on nanocomposite bioceramic, highly bioactive and biocompatible, with PVA polymer. These structures must be able to incorporate different drugs in various compartments for combined therapy, which allows to eradicate the bacterial biofilm and thus, to completely eliminate the bone infection. For antimicrobial therapy, currently there are many antimicrobial agents and combinations [1,7]. It is important to assure the maxima antimicrobial efficacy by sustained and prolonged administration in the time. Levofloxacin (LEV) is successfully used in clinical record in the treatment of bone infection due to ability to penetrate into trabecular and cortical bone, minimizing the risk of resistance selection [41,42]. Moreover, LEV exhibits a sustained release in mesoporous matrices due to the strong interaction with the silanol groups [30]. On the other hand, vancomycin (VAN) is widely used during the prophylaxis and postoperative surgery for prevention and treatment of bone infection [7]. VAN is described as a tricyclic glycopeptide antibiotic commonly used for treatment of severe infections caused by Gram-positive bacteria and especially indicated for methicillin-resistant *S. aureus* (MRSA), penicillin-resistant pneumococci, or patients allergic to penicillins and cephalosporins [43-45]. Moreover, due to its hydrophilic character exhibits sustained and prolonged release in polymeric systems [46,47]. Finally, rifampin (RIF) is an antibiofilm antibiotic, which is able to attack and destroy the *Staphylococci* in biofilm. RIF must always combined with another antibiotic because bacteria can develop resistance very rapidly when it is used as a monotherapy [48]. In this sense, this drug has been reported to present a synergy when it is administrated with other compounds such as levofloxacin and vancomycin [49].

Herein, the present study proposes a 3D multifunctional scaffold as a novel drug delivery system for treatment of bone infection and bone regeneration. This 3D system will be constituted by a mixture of *nanocomposite* MGHA and PVA containing different antimicrobial agents in different compartment to achieve different release kinetics. The 3D multifunctional scaffolds will be fabricated by rapid prototyping technique using a paste formed by aqueous mixture of calcined MGHA powder and PVA. Previous to scaffolds fabrication, both LEV and VAN will be incorporated into the mesopore structure and polymer matrix, respectively. Finally, this 3D scaffolds will be coated by a gelatin/glutaraldehyde layer containing RIF to obtain an early and fast release of this antibiofilm agent. *In vitro* degradability assays in simulated body fluid and biocompatibility assays in presence of preosteoblast have been performed in order to study the bone regeneration capability of these scaffolds. Moreover, antimicrobial tests to study the effectiveness on 3D multifunctional scaffolds against *Staphylococcus* and *Escherichia* biofilms have been also reported. Fig.1 displays the schematic design of multidrug 3D scaffold as well as the processing of fabrication.

## 2. Materials and methods

*2.1. Synthesis of mesoporous ceramic powder containing levofloxacin*

Highly mesostructured *nanocomposite* MGHA formed by mesoporous glass matrix with nanoparticles of apatite embedded inside of the matrix has been synthesized through the evaporation-induced self-assembly (EISA) method [50] using a non-ionic surfactant, Pluronic F127 (BASF) as structure directing agent, and tetraethyl orthosilicate (TEOS, 98%, Sigma–Aldrich), triethyl phosphate (TEP, 99.8%, Sigma–Aldrich), and calcium chloride ($CaCl_2·4H_2O$, 99%, Sigma–Aldrich) as $SiO_2$, $P_2O_5$, and CaO sources, respectively [27]. Briefly, in a typical synthesis, 19.5 g of F127 are dissolved in 168.6 mL of absolute ethanol (99.5%, Panreac) with 12.8 mL of 1.0 M HCl (prepared from 37% HCl, Panreac) solution and 19.4 mL of Milli-Q water. Aferwards, the appropriate amounts of tetraethyl orthosilicate (TEOS, 98%, Sigma-Aldrich), triethyl phosphate (TEP, 99.8%, Sigma-Aldrich), and calcium chloride ($CaCl_2·4H_2O$, 99%, Sigma-Aldrich) as $SiO_2$, $P_2O_5$ and CaO sources, respectively, were added in 1 h intervals under continuous stirring for 4 h at 40°C and subsequently maintained under static conditions at the same temperature overnight. The resulting sols were cast in Petri dishes (9 cm diameter) to undergo the EISA method at 30 °C. The gelation process occurred after 3 days, and the gels were aged for 7 days in the Petri dishes at 30 °C. Finally, the dried gels were removed as homogeneous and transparent membranes (several hundreds of micrometers thick) and calcined at 700 °C during 6 h to remove the surfactant, organics residue, and chloride ions. Once calcined, powder MGHA material was sieved to a size lower than 40 μm for the scaffold preparation [51]. LEV loading was carried out by impregnation method, soaking the powder MGHA material in an ethanolic solution containing 5.7 mg/ml of levofloxacin and incubated during 24 h in dark and orbital stirring conditions. The ratio powder and impregnation solution was 1g per 100 ml of dissolution. Then, the samples were filtered, gently washed with absolute ethanol, and denoted as $MG_{LEV}$.

*2.2. Fabrication of 3D-scaffolds containing levofloxacin and vancomycin by Rapid Prototyping*

3D multidrug scaffolds were prepared by rapid prototyping via direct-write assembly of precursor slurry using an EnvisionTEC GmbH 3-D Bioplotter™ device [28]. The injectable slurry is formed by mixture of powder $MG_{LEV}$ with PVA by mixing 6 g of mesoporous powder after levofloxacin impregnation with 0.9 g of an aqueous PVA solution containing 34 mg of vancomycin. The obtained scaffolds were denoted as MG$_{LEV}$-PVA$_{VAN}$. With the purpose of comparing, 3D MG-PVA scaffolds without drugs and containing just one drug were fabricated in the same conditions as follows: MG-PVA, MG$_{LEV}$-PVA, MG-PVA$_{VAN}$.

Cylindrical scaffolds of 1 cm diameter x 4 mm height were fabricated layer-by-layer by direct ink deposition over a plate at 40 ºC of temperature. The final ink with the right rheological properties was then placed into a polyethylene cartridge (with a Luer-Lock adapter) and fixed to a smooth flow tapered dispensing tip for highly viscous materials with an internal diameter of 0.6 mm (EFD-Nordson). Each layer was pre-designed showing a 90º rotation with respect to the previous. To obtain the best results, the dispensing speed and pressure were slightly modified from the starting machine parameters during the dispensing process for each scaffold. The scaffold hardening process occurred by solvent evaporation and the pieces were left to dry at 40 ºC for 3 days (see Fig.1) [28].

*2.3. Coated 3D-scaffolds with Gelatin-Glutaraldehyde containing rifampin*

For achieving an early and fast release of the antibiofilm antibiotic, an additional coating of mixture Gel-Glu containing RIF was applied by using the *dip-coating* technique onto 3D scaffolds. This biopolymer coating was chosen because is approved by the US Food and Drug Administration (FDA). As a method of introducing this antibiotic in gelatin has not been described, several test were carried out to find the best conditions and proportions of reactants for and homogeneous coating of the scaffold and an early and fast release of the drug. For this purpose, two different gelatine concentrations in water (1.2 and 2.4% (w/v)) were prepared and the RIF was dissolved in both solutions at a concentration of 0.6 mg/ml. After that, both solution were mixed with a solution 0.5% (v/v) of glutaraldehyde in stirring conditions during 1 h at 20 ºC. Gelatine was, previously, cross-linked with glutaraldehyde to reduce its solubility in water [52]. After that, 3D scaffolds were soaked in such biopolymer solutions, extracted and dried at room temperature (see Fig.1). These samples were denoted as G$_{RIF}$MG$_{LEV}$PVA$_{VAN}$.

*2.4 Characterization*

Structural, textural and chemical characterization was carried out by using different techniques. X-Ray diffraction (XRD) experiments were performed on a Philips X'Pert diffractometer MPD (Eindhoven, The Netherlands) equipped with a support for thin films (grazing incidence) and CuKα radiation (40 kV, 20 mA). Textural properties were determined by N$_2$ adsorption porosimetry using a Micromeritics ASAP2020 analyzer (Norcross, USA). Surface area was determined utilizing the multipoint Bruneauer-Emmett-Teller method included the software. Chemical composition was determined by elemental analysis (C, H, N) carried out on a LECO CHNS-932 microanalyzer (Saint Joseph, Michigan, USA) and Fourier Transformed Infrared spectroscopy (FTIR) in a Thermo Nicolet Nexus spectrophotometer (Thermo Scientific, USA) equipped with the Goldengate accessory for Attenuated Total Reflection (ATR). Microstructure of the scaffold was examined by scanning electron microscopy (SEM) using a field emission microscope JEOL JSM-6335F (Tokyo, Japan) at an acceleration voltage of 10 kV.

*2.5. In vitro degradability assay in simulated body fluid*

With the aim of evaluating the *in vitro* degradation as well as the bioactivity level of $G_{RIF}MG_{LEV}PVA_{VAN}$ 3D scaffolds were immersed at different time periods for 30 days in a total volume of 50 ml of simulated body fluid (SBF) under continuous stirring (100 rpm) [53]. After that, 3D scaffolds were removed and gently washed with Milli-Q water. The variation of calcium concentration and pH in the SBF during the experiment was performed by ion selective electrode technique in an Ilyte system ($Na^+$, $K^+$, $Ca^{2+}$, pH). The surface changes onto the 3D scaffolds during the degradability test were evaluated by scanning electron microcopy (SEM) after different times. Moreover, the changes occurred onto scaffold surface as a function of soaking time in SBF were studied by transmission electron microscopy (TEM) in a JEOL 3000 FEG electron microscope fitted with a double tilting goniometer stage (45°) and with an Oxford LINK EDS analyzer.

*2.6. In vitro Drug release Studies*

The *in vitro* release of different drugs from 3D $G_{RIF}MG_{LEV}PVA_{VAN}$ scaffolds were carried out by different techniques, previously checking the not interaction between them. Moreover, the release kinetics of each drug has been studied separately to verify the non-competitiveness between drugs. For these purposes it has been studied the release profiles in the scaffolds $MG_{LEV}PVA_{VAN}$, $MGPVA_{VAN}$, $MG_{LEV}PVA$ and $G_{RIF}$-MG-$PVA_{VAN}$. The obtained results show that the release profiles are identical alone and in presence of two or more drugs.

In this sense, the *in vitro* release was carried by soaking individually each 3D scaffold in 25 ml of PBS (Phosphate Buffer Saline). Then, they were introduced into an Incubator-Shaker, where they were smoothly stirred at 100 rpm and at 37 °C. For the LEV release, since it is a fluorescent molecule, its determination was performed using a spectrofluorimeter BiotekPowerwave XS, version 1.00.14 of the Gen5 program, with a $\lambda_{excitation}$ = 292 nm and $\lambda_{emmision}$ = 494 nm [30]. *In vitro* VAN release over time was determined by High Performance Liquid Chromatography (HPLC) using a 1260 Infinity HPLC system (Agilent) and UV-Vis analysis at 290 nm at different times [46]. It has been proved that no interferences appear between VAN and LEV in HPLC. Finally, to determine the *in vitro* RIF release UV-Vis spectroscopy technique was used, since it exhibits a band at 474 nm in UV-Vis which was chosen because it does not interfere with the band of VAN. Determination of the concentration of RIF over time was performed by using Helios Zeta UV-VIS spectrophotometer at 474 nm at different times [54].

*2.7. Antimicrobial activity of 3D scaffolds*

Prior to the bacterial tests, the samples were sterilized by UV-light radiation during 10 min in both sides. The effectiveness of these multidrug 3D systems against *Staphylococci* in biofilm has been determined. Previously, biofilms of *S. aureus* was developed and incubated onto cover glass disks. For this purpose, cover glass disks was suspended in a bacteria solution of $10^8$ bacteria per ml during 48 h at 37 ºC and orbital stirring at 100 rpm. In this case the medium used was 66% TSB + 0.2% glucose medium to promote robust biofilm formation. After that, the cover glass disks containing biofilm were localized onto six well culture plates (CULTEK) in 6 ml of new medium. Then, multidrug 3D scaffolds were submerged avoiding the direct contact with biofilm coated glass disk. After different times of incubation, the glass-disk were washed three times with sterile PBS, stained with a 3 μl/ml of Live/Dead® Bacterial Viability Kit (BacklightTM) and 5 μm/ml calcofluor solution to specifically determine the biofilm formation, staining the mucopolysaccharides of the biofilm (extracellular matrix in blue) [55]. Both reactants were incubated for 15 min at room temperature. Biofilm formation was examined using a in an Olympus FV1200 confocal microscope. In order to check the strategy proposed by using multidrug systems, scaffolds only containing only VAN and LEV and both were also evaluated in order to evaluate the effectiveness of combined therapy.

Parallel, the effectiveness of these 3D multidrug systems have been also carried out against gram negative bacteria as *Escherichia coli* (*E. coli*). Similar experiments were carried out for E. coli biofilm. Previously, *E. coli* biofilms were formed onto cover glass disks by inoculated $10^8$ per ml bacteria solution during 48 h at 37 ºC and orbital stirring at 100 rpm. After different times of incubation, the glass-disk were washed three times with sterile PBS, stained with Live/Dead® Bacterial Viability Kit (BacklightTM) and calcofluor in similar way that *S. aureus* test and examined in confocal microscope.

In addition, the determination of number of colony-forming units (CFU) present in each biofilm (Gram-positive and gram-negative) before and after treatment with 3D multidrug systems was carried out. Each biofilm coated glass disk was previously sonicated during 15 minutes in 6 ml of sterile PBS to remove all bacteria from biofilm. After, that 100 µl of the bacteria suspension was cultivated on Tryptic Soy Agar (TSA) (Sigma Aldrich, USA) plates, followed by incubation at 37 ºC overnight. Then, CFUs were counted and represented as Log [CFU]. Six replicated were done for each 3D multidrug system at different times (1, 6 and 24 h).

2.8. In vitro biocompatibility studies: MC3T3-E1 preosteoblast culture

Previous to *in vitro* assays with osteoblast cells, all samples were sterilized by UV radiation during 10 min. Cytotoxicity, cell morphology, mitochondrial activity and cell differentiation studies were carried out by utilizing MC3T3-E1 pre-osteoblasts (mouse osteoblastic cells able to differentiate to osteoblast or osteocytes) cultured on the 3D-scaffolds of different compositions. For this purpose, the cells were cultured in complete medium Eagle medium alpha modified by Dulbecco ($\alpha$ DMEM) supplemented with 2 mM glutamine, 100 U ml$^{-1}$ penicillin, 100 g·ml$^{-1}$ streptomycin and fetal bovine serum (FBS) at 10% at 37 ºC under atmosphere conditions of 95% humidity and 5% $CO_2$. MC3T3-E1 cells in a concentration of $2.5·10^5$ cells/mL in complete medium were cultured on the 3D-scaffolds previously placed in 24-well plates and cultured in 5% $CO_2$ and 95% humidity atmosphere at 37 ºC during different times. Control samples corresponds to empty positions where the same quantity of cells was cultured.

*Citotoxicity Test – LDH*

Activity of the lactate dehydrogenase enzyme (LDH) was determined in the culture medium in contact with the scaffolds after 1 and 7 days of incubation. Activity of LDH released by the MC3T3-E1 cells is directly related to the rupture of the plasmatic membrane (cell death) that, when broken, releases all organelles and enzymes present in the cytoplasm. Measurements were performed by using a commercial kit (Spinreact) at 340 nm with a UV-Visible spectrophotometer Unicam UV-500.

*Cell morphology studies – SEM*

For these studies, adherent cells of different samples were fixed with glutaraldehyde (2.5% in buffer solution, PBS) for 45 min. Afterwards, all the samples were gradually dehydrated through the progressive replacement of water with a serie of ethanol solutions (30%, 50%, 70% and 90%) for 30 min with a final dehydration in absolute ethanol for 60 min. Later, samples were introduced into a vacuum oven at 40°C for 7days. After this time, samples were stick onto cupper holder and finally metallized with gold in a metallization device EMS150R-S. Study of the cell morphology on different scaffolds was performed by using SEM in a JEOL 3565F.

*Cell morphology and colonization studies – Confocal microscope*

Fluorescence microscopy was performed with a confocal laser scanning microscope OLYMPUS FV1200 (OLYMPUS, Tokyo, Japan), using a 60x FLUOR water dipping lens (NA=1.0). The images were prepared for analysis using Software 3D Imaris to project a single 2D image from the multiple Z sections by using an algorithm that displays the maximum value of the pixel of each Z slice of 1 μm of depth. The resulting projection was then converted to a TIF file using this software. In the images, DAPI and Atto 565–phalloidin were visualized in blue and red, respectively. The reflection of the scaffold material was visualized in green. The dyeing process was performed using the methodology described elsewhere [56].

*Mitochondrial Activity – MTT*

For evaluating cell mitochondrial activity of living cells on the different scaffolds as well as its surroundings after 1 an 7 days of incubation the MTT method was employed. This method is based in the reduction of 3-(4,5-dimethylazol-2-yl)-2,5-diphenyltetrazolium (yellow) to blue formazan. This measurement was used in terms of cell proliferation as described in previous works. For this purpose, culture medias were substituted with 1 ml of DMEM and 125 μL of 0.012 g·ml$^{-1}$ MTT solution in PBS. Samples were incubated for 4 h at 37°C and 5% $CO_2$ in dark conditions. Then, media was removed and 500 μL of HCl-isopropanol solution 0.4 M. Finally, absorbance at 570 nm was measured by using a Helios Zeta UV-VIS spectrophotometer.

*Alkaline Phosphatase Activity – ALP*

ALP of the cells growing onto the scaffolds was utilized as marker of cellular differentiation in the evaluation of the phenotype expression of osteoblasts. ALP was measured employing the Reddi-Huggins method based in the hydrolysis of p-nitrophenylphosphate to p-nitrophenol. For this purpose, MTC3T3-E1 pre-osteoblastic cells (2.5·10$^5$ cells·ml$^{-1}$) were cultured directly on top of the scaffolds in a 24-well plate and incubated under standard culture conditions using media supplemented with β-glycerolphosphate (50 mg·ml$^{-1}$) and L-ascorbic acid (10 mM). For evaluating ALP of both, scaffolds and its surroundings, after 7 days of incubation, each scaffold was transferred to a new well. The measurements have been normalized with respect the amount of protein total determined by Bradford colorimetric method.

*Mineralization test*

Matrix mineralization was measured by alizarin red staining after cell incubation with G$_{RIF}$MG$_{LEV}$PVA$_{VAN}$ scaffolds for 10 days, as described [57]. Since the samples contain calcium in its composition this study has been conducted on the well. Stain was dissolved with 10% cetylpyridinum chloride in 10 mM sodium phosphate, pH 7, and measuring absorbance at 620 nm. Moreover, the mineralization process has been studied by a deep surface characterization with XRD with and also without presence of preosteoblast cells. The high surface roughness of the samples, when compared with a standard powder XRD preparation, has hampered the XRD measures giving rise to a poor signal-to-noise ratio.

*Statistical Analysis*

Data was expressed as average/mean ± standard deviation in three experiments. Statistic analysis was performed by using the software Statistical Package for the Social Sciences (SPSS) version 11.5. Statistical comparatives were carried out through variance analysis (ANOVA). Scheff proof was utilized for the post hoc evaluation of the differences among groups. In all statistical evaluations, *p<0.05* was considered as statistically significant.

## 3. Results and Discussion

All scaffolds were structural, chemical and morphological characterized by different techniques. Firstly, the structural characterization of the MGHA powder (before and after LEV loading) and different 3D scaffolds (MGPVA, MG$_{LEV}$PVA, MGPVA$_{VAN}$, MG$_{LEV}$PVA$_{VAN}$ and G$_{RIF}$MG$_{LEV}$PVA$_{VAN}$) was carried out by XRD to observe both, its mesoporous arrangement and the presence of nanocrystalline hydroxyapatite embedded into glassy matrix, respectively. Fig.2 displays XRD patterns corresponding to MGHA$_{LEV}$ powder and 3D scaffolds before and after coating (MG$_{LEV}$PVA$_{VAN}$ and G$_{RIF}$MG$_{LEV}$PVA$_{VAN}$). Low angle XRD pattern (left) shows a well-defined diffraction maximum at $2\theta = 0.86$ degree and wide maxima around $2\theta = 1.43$ and $1.67$ degree, which could be indexed as 10, 11, and 20 reflections of a 2D-hexagonal structure with *p6mm* plane group, based on TEM study (see below). Wide angle XRD pattern (right) reveals the presence of nanocrystalline apatite phase exhibiting (002), (211) and (310) reflections. These results highlight the fact that both prototyping scaffolding and posterior dip-coated processes do not affect to the structural order of the material, maintaining the 2D-hexagonal structure and nanocrystallinity of the apatite phase present in them. Moreover, the presence of different antibiotics on the scaffolds do also not affect to the mesoporous structure and nanocrystallinity of the nanocomposite mesoporous material.

Textural properties revealed a significant decrease in both, specific surface area and total pore volume, before and after the LEV drug loading, respectively. It is reported in the literature that the decreasing in the values of specific surface area, total pore volume and pore diameter after drug loading process is related to the confinement of the drug into the mesoporous structure [58]. Table 1 shows the percentages (%) of each antibiotic present in each scaffold with respect of the weight of the scaffold and the variation of their textural properties after impregnation process, conformed by RP technique and dip-coated process, respectively.

Concerning to the yielding of the drug loading process, it was determined by combination of elemental chemical analyses and TG studies. The percent of nitrogen, carbon and hydrogen for all samples obtained by elemental analysis have been collected in the Table S1 in the supporting information. Taking into account that both MGHA powder and PVA polymer do not content nitrogen in the composition, then LEV and VAN were determined by calculation of the % of nitrogen in the sample. Concerning LEV amount, the % of nitrogen of MGHA powder after impregnation process and scaffolding remains unchanged with a value of 3%. Although a priori, this percentage would seem low, it is within the normal range of loading of a drug by the impregnation method [59]. The amount of VAN was determined in presence or not of VAN obtaining a value of 1.8 %. Finally, the amount of rifampin was determined by two different methods by TG analyses (data not shown) on the coating scraped from the scaffolds and in presence of not of antibiotic by showing a percent of 2.5 %. This percent was confirmed by an indirect method, by total dissolution of RIF coating in acidic medium and determination by UV-Vis at 474 nm.

Morphology study of the different 3D scaffolds was performed by SEM. Before dip-coating process, the MG$_{LEV}$PVA$_{VAN}$ scaffolds exhibit a very high porosity in both, surface and inner structure. Giant macropores of about 1 mm can be observed as well as a regular and high porosity all over and in the scaffold (Fig.3A). High magnification shows the presence of mesopores of around 50 µm (Fig.3B) and a similar view is observed in the cross-section (CS) (see supporting information Fig.S1), which indicated the high rate of interconnectivity. A detail of the surface shows a smooth polymeric surface with incrustations of MGHA ceramic material (Fig.3C and 3D). It is important to keep in mind that free drug 3D-scaffolds exhibit similar morphological features, showing that the incorporation of different antibiotics does not affect to the morphology of the scaffold. In addition, TEM image and FT diagram (Fig.3E) display a mesoporous arrangement in the 2D hexagonal structure, showing the 3D hierarchical structure of MG$_{LEV}$PVA$_{VAN}$ scaffold.

In order to obtain an early release of RIF antibiofilm, MG$_{LEV}$PVA$_{VAN}$ were coated by dip-coating method with a gelatin-glutaraldehyde mixture containing a RIF solution. In this sense, this external coating should, once delivered, break the *biofilm* capsule allowing the VAN and LEV activity. SEM studies the 3D scaffolds after coating containing rifampin are shown in Fig.4. The obtained results show that after coating the macropores corresponding to 1 mm are obstructed by a layer of thickness of 5 µm.

*In vitro* degradability assay in SBF of G$_{RIF}$MG$_{LEV}$PVA$_{VAN}$ was performed at different key times (Fig.5). After 1 h of incubation, SEM image displays the total dissolution of the coating leaving empty pores 1 mm. In addition, the *in vitro* degradability assays in SBF were performed during long periods of incubation at 15 and 30 days. The obtained results show a partial degradation of scaffold with the appearance of pores on the scaffolds surfaces. Higher magnification micrographs show also a typical layer of hydroxyapatite formed, which is identified by the needle-like particles observed all over the surface of the scaffolds indicating a high level of bioactivity. Parameters such as pH and [Ca$^{2+}$] were also monitored during this experiment, showing as initial increasing of calcium content until 7 days followed of a decreasing until 30 days which indicates the bioactivity process (see supporting information Fig.S2). To confirm this bioactive process, TEM studies of the scrapped surfaces were carried out and collected in Fig. 6. The obtained results confirm a formation of a nanocrystalline hydroxyapatite layer onto the surfaces of 3D multifunctional scaffold after 7 days in SBF.

*In vitro* release tests of different 3D scaffolds containing drugs were carried out in PBS at pH 7.4 and 37 ºC. Fig.7 shows the different LEV, VAN and RIF release profiles from the G$_{RIF}$MG$_{LEV}$PVA$_{VAN}$ scaffolds. In the case of VAN and LEV, the release profiles are more sustained and prolonged in time in comparison with the RIF release, which is characterized by a fast release in just one hour, according to *in vitro* degradability tests. Fig.S3 displays the dosages corresponding to G$_{RIF}$MG$_{LEV}$PVA$_{VAN}$ sample after 1, 6 and 24 h of incubation. The minimum inhibitory concentration (MIC) of each antibiotic is 0.5 µg/ml [60,61] for RIF and 0.5-1 µg/ml [62] for LEV and VAN. Since the release dosage from $G_{RIF}MG_{LEV}PVA_{VAN}$ scaffolds displays an initial effective dosage formed by three antibiotics followed of an effective and sustained dosage for long periods of time (> 10 days) for LEV and VAN.

For porous and polymer matrices, it has been reported that drug release can be mediated by diffusion, erosion/degradation and swelling followed by diffusion. Although some matrix degradation is involved, under perfect sink conditions, the main driving force for drug delivery out of the mesoporous matrices is pore diffusion/convection, which can be fitted to first order kinetics. Additional parameters such as drug-carrier and host-guest interactions are key factors to dictate drug release profiles. Drug molecules could directly interact with the matrix, in this case, the mesoporous material, the PVA polymer and the Gel-Glu mixture retarding their release. The association and dissociation processes are assumed to be reversible. Furthermore, in general, the reversible association of a drug molecule with a carrier is assumed to follow first order kinetics. Therefore, the theoretical model adopted in this work considers first-order diffusion/convection and drug association/dissociation. Concretely, drug release patterns correspond to fast diffusion/convection but slow association/dissociation. This leads to a decoupling of drug association/dissociation from drug diffusion/convection: the fast release of initially free drug molecules via diffusion/convection and the slow release of initially bound drug molecules that is dictated by the dissociation process. Accordingly, two first order kinetics or two exponential release mechanisms can be described as follows [63].

$$\frac{Q_t}{Q_0} = \frac{k_{off}}{k_{on}+k_{off}}\left(1 - e^{-k_s t}\right) + \frac{k_{on}}{k_{on}+k_{off}}\left(1 - e^{-k_{off} t}\right)$$

Eq: 1

Where $Q_t$ is the cumulative drug release a time t; $Q_0$ is the initial amount of drug; $k_s$ is the rate constant of diffusion/convection; and $k_{on}$ and $k_{off}$ are the rate constants of association and dissociation respectively. The free energy difference between the free and bound states, ΔG, determines the amounts of initially free and bound drug and can be calculated by the following equation 2:

$$\Delta G = -k_\mathrm{B} T \ln\left(\frac{k_\mathrm{on}}{k_\mathrm{off}}\right)$$

Eq: 2

Where $k_B$ is Boltzmann's constant and T is the absolute temperature (assumed to be 310K). In this study, therefore, three parameters, $\Delta G$ (instead of $k_{on}$), $k_s$ and $k_{off}$ were used to describe the cumulative drug release from the fabricated scaffolds. Fitting experimental release patterns to eq. 1 allowed the determination of the experimental values for $k_s$, $k_{on}$ and $k_{off}$. Then, $\Delta G$ was calculated using eq. 2. The experimental results are summarized in Table 2.

Drug release profiles can be classified into four distinct categories (A-D). The categories are based on the magnitude of the initial burst release, the extent of the release and the steady-state release kinetics folowing the burst release. The burst release is described as the initial release before reaching the steady-state. In types A and B, the initial burst release is followed by little additional release. The release profile in type C is similar to that in type A in terms of initial burst, but the burst release is followed by a steady-state release of the remaining drug. Finally, type D exhibits a low initial burst release followed by a steady-state release until most of the loaded drug is released [64,65].

From the data of the Table 2, kinetic profiles of the drugs can be classified. In the case of LEV release, $k_s \gg k_{off}$ wich indicates that difussion and convection are not neglected during the steady-state release. Accordingly, LEV release profile can be classified as type B, exhibiting a low initial burst release with steady-state release.

RIF also presents $k_s \gg k_{off}$, however, value of $\Delta G$ is higher, meaning rifampicin release profile belongs to type C kinetic release. Finally, VAN presents $k_s < k_{off}$ with a very high value of $\Delta G$, which indicates a high burst effect at the beginning of the release but a high interaction between the drug and the matrix (PVA) meaning a steady-state release corresponding to type A.

Once determined release profiles, antimicrobial assays have been carried out to determine the effectiveness of these multidrug systems against *S. aureus* biofilms. These antimicrobial studies has been carried out for all scaffold containing one, or two or three drugs in order to determine the success of the proposed combined therapy. Fig.8 displays the most representative results derived of the effects of MG$_{LEV}$PVA$_{VAN}$ and G$_{RIF}$MG$_{LEV}$PVA$_{VAN}$ 3D scaffolds onto the preformed *S. aureus* biofilm by confocal laser scanning microscopy (CLSM). Initially, the preformed biofilm shows a typical structure formed by colony live bacteria (green) covered by protective mucopolysaccharide matrix (blue). After 1 h of incubation with the different samples, notable differences are observed in presence or not of RIF. MG$_{LEV}$PVA$_{VAN}$ containing both LEV and VAN is not able to destroy completely the biofilm, appearing even live bacteria colonies coated with its protective layer as it can observe in Fig.8. However, G$_{RIF}$MG$_{LEV}$PVA$_{VAN}$ scaffolds totally destroy the biofilm, observing colonial killed bacteria without the presence of protective layer of mucopolysaccharides. These results agree with *in vitro* drug kinetic assays (Fig.7), where a sudden release of all amount of RIF present on the scaffold was released together with a sustained release of both VAN and LEV in the first hour of incubation (Fig.S3), which it seems to be an effective strategy for the elimination of bacteria biofilm. After 24 h, the G$_{RIF}$MG$_{LEV}$PVA$_{VAN}$ scaffold shows complete destruction of biofilm appearing isolated fragments, while MG$_{LEV}$PVA$_{VAN}$ scaffolds still exhibits bacteria colonies with protective covered, indicating no efficacy against biofilm. It is important to remark that after long time of exposure all scaffold containing one or two drugs showed a total destruction of studied biofilm. These results show that our multidrug systems formed by the combination of RIF, LEV and VAN together with the designed strategy designed are very effective for the total destruction of biofilm in the first 24 h of incubation, which is indicative of their antimicrobial efficient [66].

In general the resulting biofilm in a common infection case may not be single species biofilm. Thus, the effectiveness of these multidrug systems has been also determined on gram-negative pathogenic as E. coli. Fig.9 shows the most significant results of *E. coli* biofilm after different times of incubation with G$_{RIF}$MG$_{LEV}$PVA$_{VAN}$. At it can be observed, initially the effectiveness against Gram-negative biofilm is lower with respect to Gram-positive, due to after 1 h of incubation still appear small colonies of 20 μm, which are formed by live bacteria (green) coated with a protective matrix (blue). After 6 and 24 h of incubation, the scenary is very similar to *S. aureus* a few scattering formed by small fragments of protective biolayer (blue), dead bacteria (red) and small amount of lived bacteria (green).

In order to quantify the number of live bacteria present in each biofilm after of different treatments, CFU was determined. The results were summarized in the Fig.10. The obtained results shows notable differences in both 3D scaffolds, containing and not RIF. For MG$_{LEV}$PVA$_{VAN}$, the histogram evidences bacterial survival above $10^2$ CFU after 24 h of incubation in both pathogens. It has been reported that a few live bacteria as 10-100 CFU can cause an infection, which show the inefficient of this system [67]. On the contrary, the CFU analysis for the G$_{RIF}$MG$_{LEV}$PVA$_{VAN}$ systems display a notable efficiency against both Gram-negative and Gram-positive bacteria, showing values below 10-100 CFU after 6 and 24 h of incubation, respectively.

With the purpose of evaluating the biocompatibility of the fabricated 3D scaffolds for achieving bone regeneration, *in vitro* studies with MC3T3-E1 preosteoblast cells were carried out. The studies were performed on all scaffold (MGPVA, MG$_{LEV}$PVA$_V$ and G$_{RIF}$-MG$_{LEV}$PVA$_{VAN}$, to determine the influence of different component in the biocompatibility studies However, the most representative assays onto G$_{RIF}$MG$_{LEV}$PVA$_{VAN}$ are shown in this manuscript. Parameters such as cytotoxicity (related to LDH), proliferation (related to mitochondrial activity), cell morphology and cellular differentiation (related to ALP) were studied (Fig.11 and Fig.S4). As it can be observed in Fig.11, in LDH assay, control well and G$_{RIF}$MG$_{LEV}$PVA$_{VAN}$ scaffold do not present significant differences, which indicates a good level of biocompatibility. This fact reveals the non-delivery of "cytotoxic products" from the scaffold to the cell media and also that, the amount of delivered drug in this time (1 day) has not a cytotoxic effect for the cells and its surroundings (Fig.S4). Regarding the proliferation studies after 7 and 15 days (MTT), results show an adequate proliferation on the scaffold and no significant differences observed with respect to the control well indicating a good cell colonization and proliferation on the scaffolds. Cellular differentiation studies at 7 days show a slight decreasing of ALP activity could be attributed to the presence of antibiotics in the media and it is an indicator of the influence of the antibiotics on the differentiation process. However, after 15 days (Fig.S4) any significant differences are observed in all scaffolds, which could be indicative of the initiation of remodeling process observed after *in vitro* studies in SBF (Fig.5 and Fig.6). Finally, mineralization studies have been carried out by alizarin test onto well plate and a deeper XRD study on the scaffold surfaces after 10 days in in vitro culture with and also without presence of preosteoblast cells. The obtained results are collected in Fig.12, showing no significant differences on the alizarin test between G$_{RIF}$MG$_{LEV}$PVA$_{VAN}$ scaffold and control. However, the XRD studies clearly show the mineralization process by formation of a nanocrystalline apatite phase on the surface of these scaffolds, whose crystallinity increases in the presence of preosteoblasts.

Cell morphology and cell colonization were studied by SEM and CLSM. Fig.13A shows a SEM micrograph corresponding to the G$_{RIF}$MG$_{LEV}$PVA$_{VAN}$ scaffolds after 7 days of incubation showing extended cells in entered surface of scaffolds emitting filopodia as communication and anchoring elements. The cells colonizing all surface of scaffolds were visualized by staining (Fig.13B). Active cytoskeleton was stained with phalloidin and the nuclei with DAPI and cells were observed by confocal laser scanning microscopy. The image indicates that, after 7 days of culture, MC3T3-E1 cells have proliferated onto this scaffold, in agreement with the image obtained by SEM. In order to visualize those cells migrating upward, 200 sections of 1 μm thickness along Z axis were acquired and processed in single 2D images. In this case, the image shows an aggrupation of spread cells colonizing the pore walls of the scaffold, showing cells well-developed with actine cytoskeleton organized into long parallel bundles, extending protrusions in the direction of migration.

**Conclusions**

A novel therapeutic solution for bone infection treatment based on 3D multifunctional scaffolds, which combines the merits of osseous regeneration and local multidrug delivery has been developed. The 3D multidrug scaffolds, containing rifampin, levofloxacin and vancomycin, have been designed by rapid prototyping of mixture of nanocomposite bioceramic and polyvinyl alcohol with an external coating of gelatin-glutaraldehyde. The different antimicrobial agents have been localized in different compartments to achieve different release kinetics. These 3D multidrug scaffolds, exhibiting an early and fast release of rifampin followed by sustained and prolonged release of vancomycin and levofloxacin, are able to destroy *Staphylococcus* and *Escherichia* biofilms as well as inhibit bacteria growth in very short time periods. This new combined therapy approach involving the sequential delivery of antibiofilms with antibiotics constitutes an excellent and promising alternative for bone infection treatment.

**Acknowledgments**

MVR acknowledges funding from the European Research Council (Advanced Grant VERDI; ERC-2015-AdG Proposal No. 694160). The author also thanks to Spanish MINECO (CSO2010-11384-E, MAT2015-64831-R and MAT2013-43299-R). The authors wish to thank the ICTS Centro Nacional de Microscopia Electrónica (Spain),

CAI X-ray Diffraction, CAI NMR, CAI Cytometer and Fluorescence microscopy of the Universidad Complutense de Madrid (Spain) for the assistance. RGA was supported by European Commission (EACEA) through MONABIPHOT Master Course.

**Tables:**

**Table 1:** Data corresponding to the textural properties as surface area ($S_{BET}$), pore volume ($V_p$) and pore diameter ($D_p$). Percent of PVA and different antimicrobial agents presents in the different scaffolds. In the table is also shown the MGHA powder and conformed by rapid prototyping with PVA.

| SAMPLE | $S_{BET}$ (m$^2$·g$^{-1}$) | $V_p$ (cm$^3$·g$^{-1}$) | $D_p$ (nm) | % PVA* | % LEV* | % VAN* | % RIF** |
|---|---|---|---|---|---|---|---|
| **Powder MGHA** | 178.0 | 0.35 | 8.5 | - | - | - | |
| **MG-PVA** | 145.5 | 0.30 | 8.7 | 20.6 | - | - | |
| **MG$_{LEV}$-PVA$_{VAN}$** | 27.6 | 0.05 | 4.5 | 19.0 | 3 | 1.8 | - |
| **G$_{RIF}$-MG$_{LEV}$-PVA$_{VAN}$** | 19.4 | 0.04 | 7 | 23.1 | 3 | 1.8 | 2.5 |

(*) Determined by elemental analyses of Table S1; (**) Determined by combination of TG study and indirect method.

**Table 2:** Release kinetic parameters of rifampin, levofloxacin and vancomyci delivery from different matrices

| Drug release | $\Delta G$ (x10$^{-21}$J) | Ks (h$^{-1}$) | Koff (h$^{-1}$) | |
|---|---|---|---|---|
| **LEV** | **1.7** | **0.13** | **0.0006** | **G$_{RIF}$-MG$_{LEV}$-PVA$_{VAN}$** |
| **LEV** | 1.9 | 0.14 | 0.0005 | **MG$_{LEV}$-PVA$_{VAN}$** |
| **LEV** | 1.8 | 0.14 | 0.0006 | **MG$_{LEV}$-PVA** |
| **VAN** | **7.5** | **0.02** | **0.89816** | **G$_{RIF}$-MG$_{LEV}$-PVA$_{VAN}$** |
| **VAN** | 8.0 | 0.017 | 0.87937 | **MG$_{LEV}$-PVA$_{VAN}$** |
| **VAN** | 7.4 | 0.019 | 0.86973 | **MG-PVA$_{VAN}$** |
| **RIF** | **4.2** | **29.285** | **3.02853** | **G$_{RIF}$-MG$_{LEV}$-PVA$_{VAN}$** |
| **RIF** | 4.1 | 29.328 | 3.01777 | **MG$_{LEV}$-PVA$_{VAN}$** |
| **RIF** | 4.1 | 28.358 | 3.02842 | **MG-PVA$_{VAN}$** |

**Figure and Figure captions**

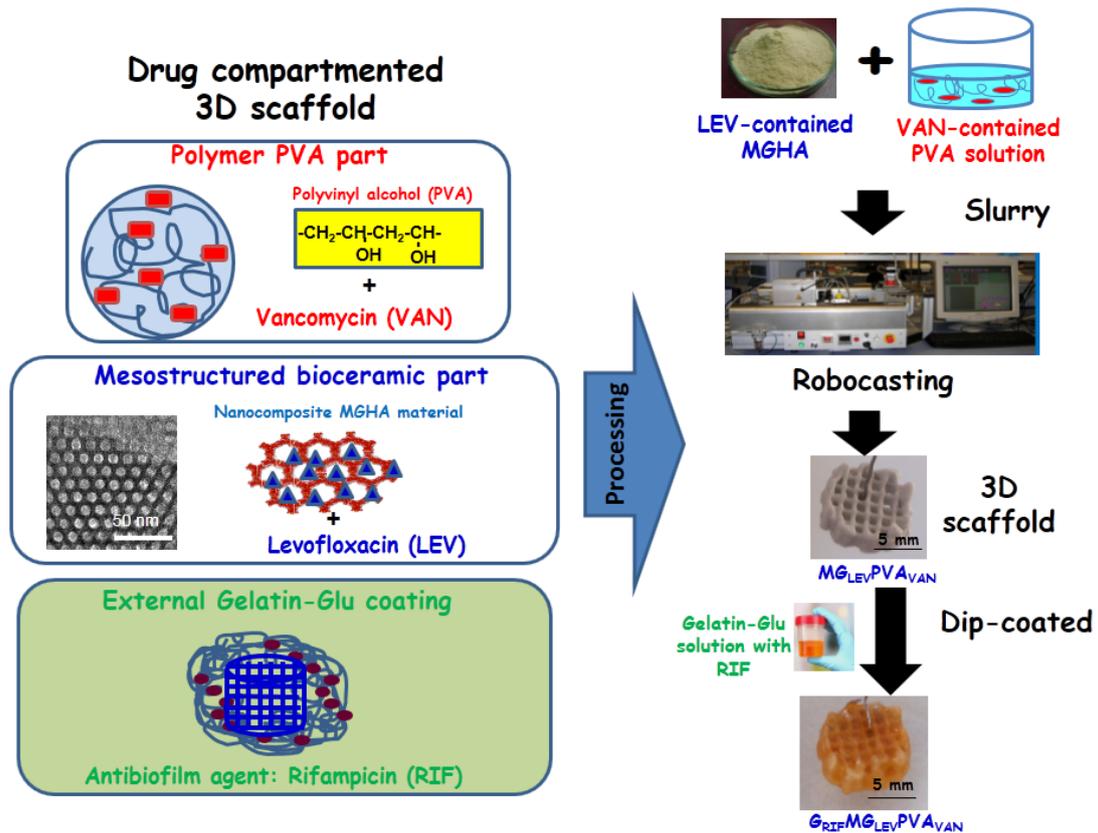

**Fig.1** Schematic design of the present study based on the development of a novel therapeutic multidrug 3D scaffolds for bone infection. These 3D scaffolds containing rifampin, levofloxacin and vancomycin, is formed by a mixture of MGHA and PVA processed rapid prototyping technique ($MG_{LEV}PVA_{VAN}$) and posterior coating with gelatin-glutaraldehyde layer ($G_{RIF}MG_{LEV}PVA_{VAN}$). (Left) schematic representation of the localization of different antibiotics in different compartment of 3D scaffold to obtain a sequential and effective release kinetics of multi-therapy. (Right) Processing of fabrication of these multifunctional 3D scaffolds for bone infection therapy.

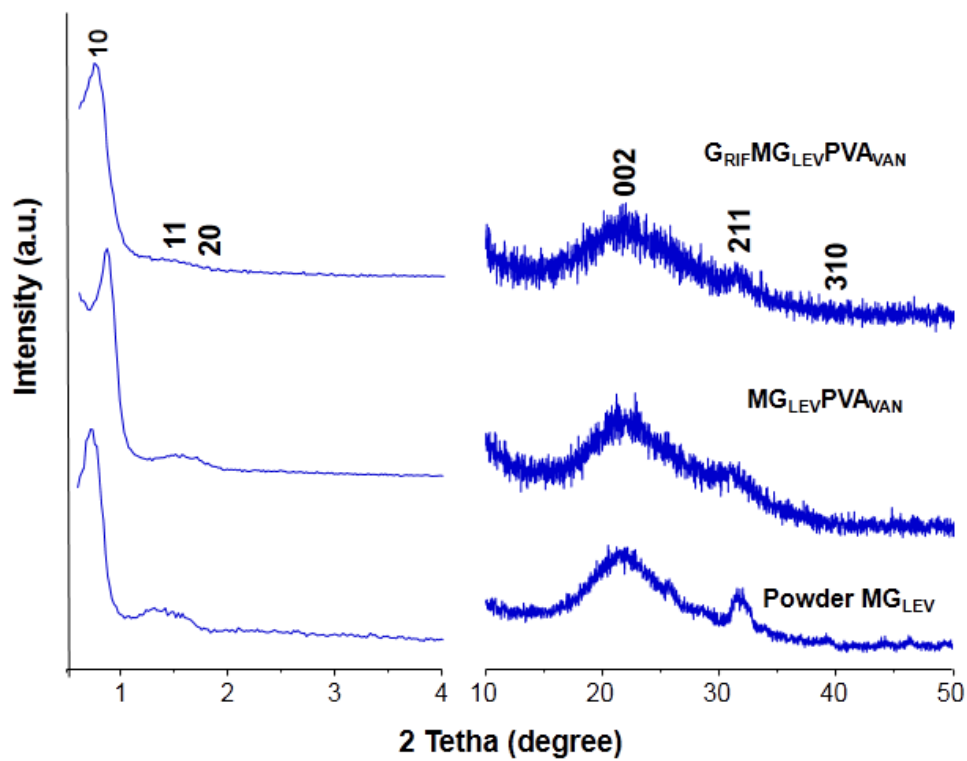

**Fig.2** XRD patterns corresponding to low (left) and wide (right) scattering angles.

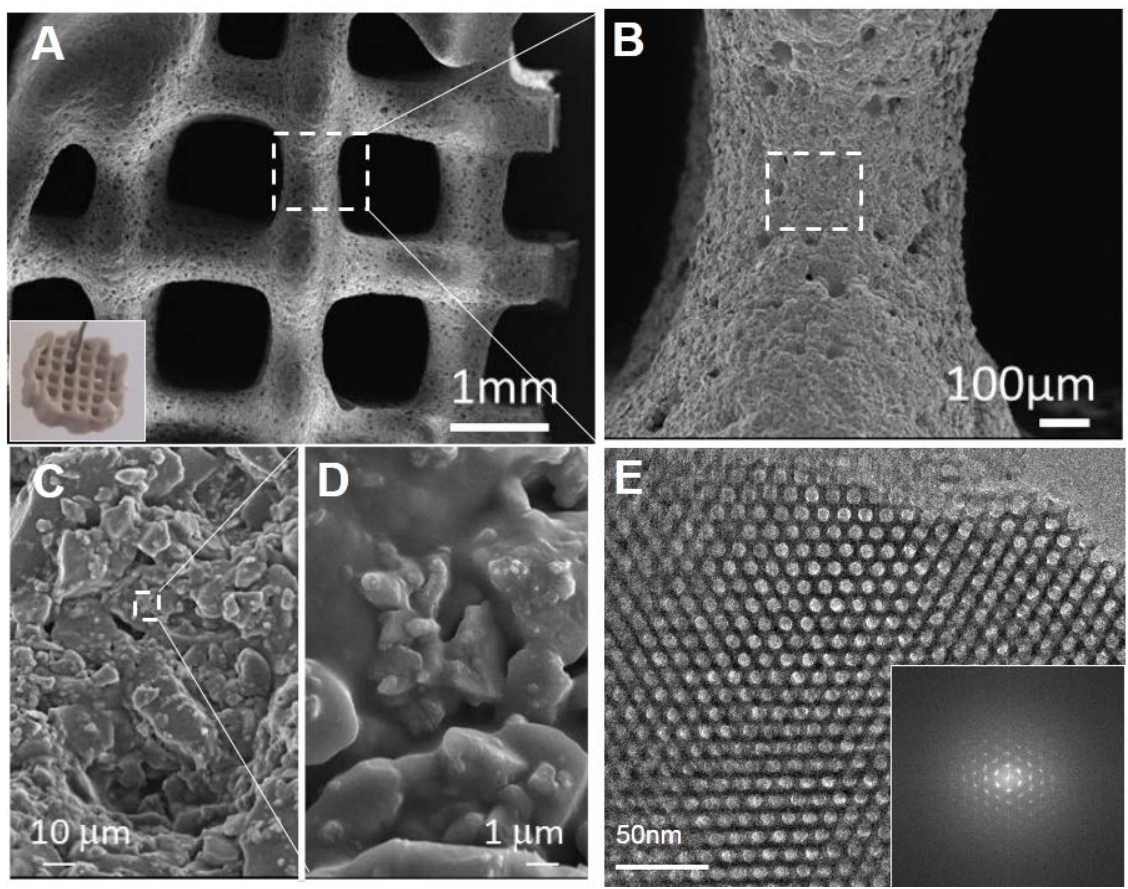

**Fig.3** Morphological and structural characterization 3D composite scaffolds MG$_{LEV}$PVA$_{VAN}$ showing a high a regular level of hierarchical porosity from macro to mesoporous scale. (A, B, C and D) SEM micrographs at different magnification. (E) TEM image and FT diagram showing the mesoporous arrangement in a 2D hexagonal structure.

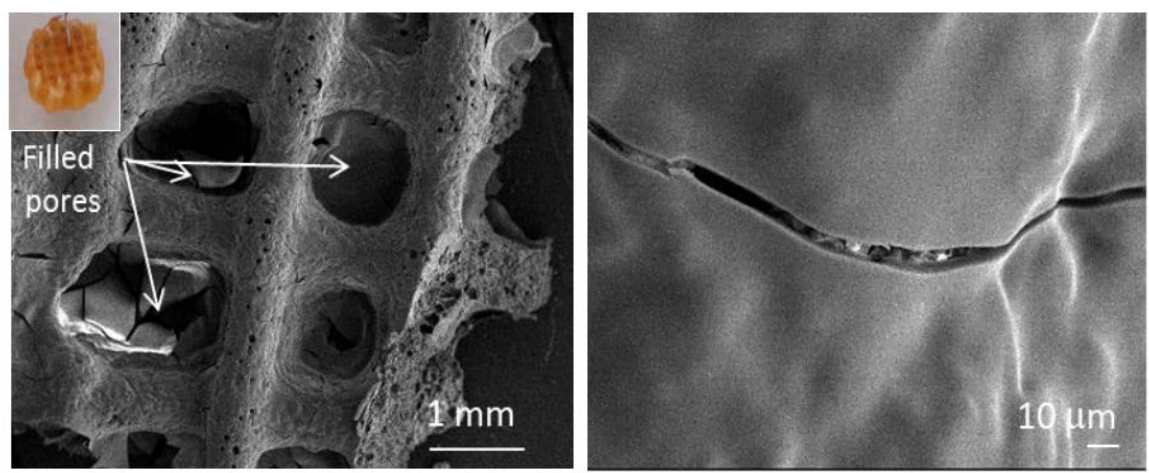

**Fig.4** SEM micrographs at different magnification corresponding to G$_{RIF}$MG$_{LEV}$PVA$_{VAN}$ 3D scaffold.

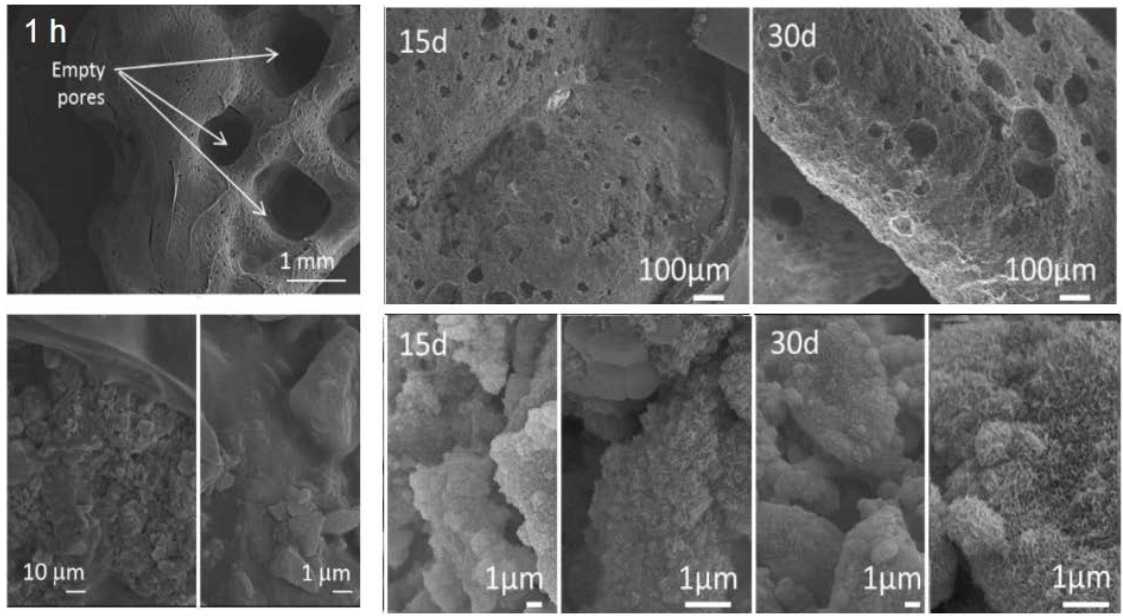

**Fig.5** *In vitro* degradability assay in SBF corresponding to G$_{RIF}$MG$_{LEV}$PVA$_{VAN}$ 3D scaffold. SEM images at different magnification showing the surface of the 3D scaffolds, after 1 h, 15 and 30 days of incubation in SBF.

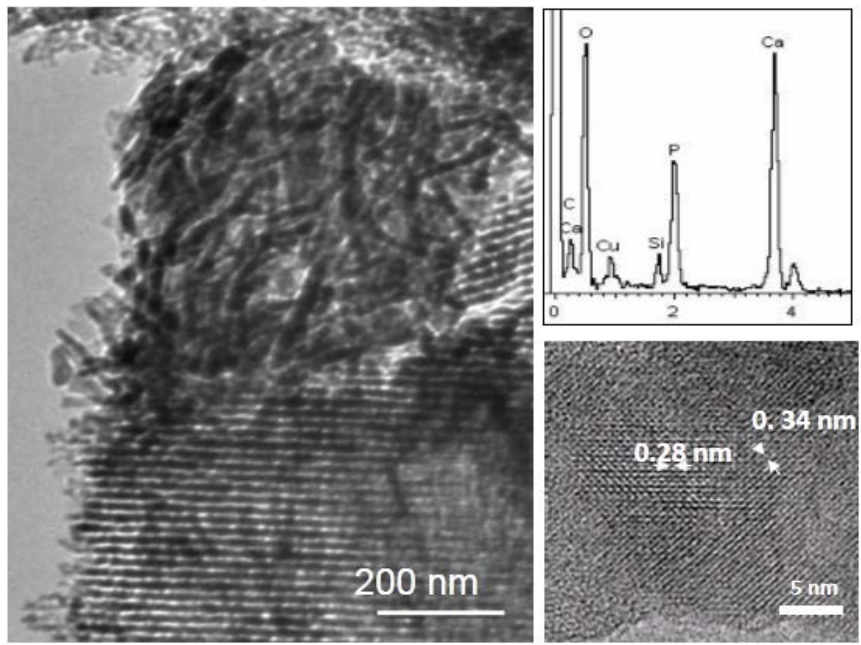

**Fig.6** TEM images of G$_{RIF}$MG$_{LEV}$PVA$_{VAN}$ after 7 days in SBF. Needle-shaped crystallites are observed together with mesoporous channels of the scaffolds. High-magnification image evidence ordered planes corresponding to (211) and (002) d-spacings of an apatite phase. EDS analyses shows that needle-shaped is Ca-deficient apatite with a Ca/P ratio of 1.60.

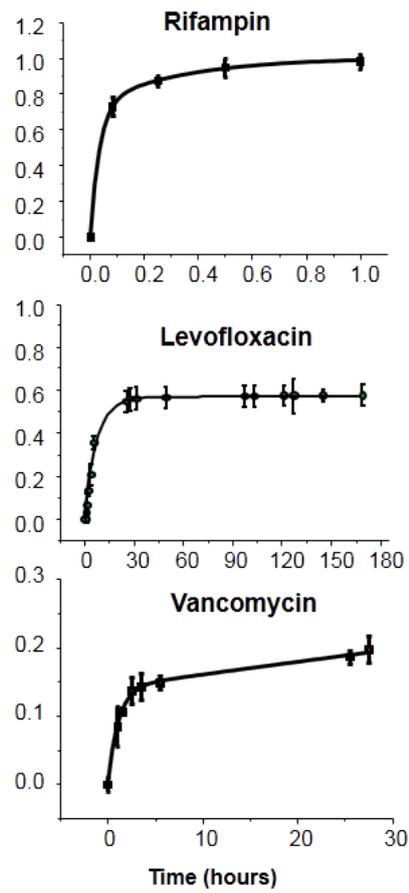

**Fig.7** Graphic representation of the RIF, LEV and VAN release kinetics fitting to the model corresponding to eq. 1.

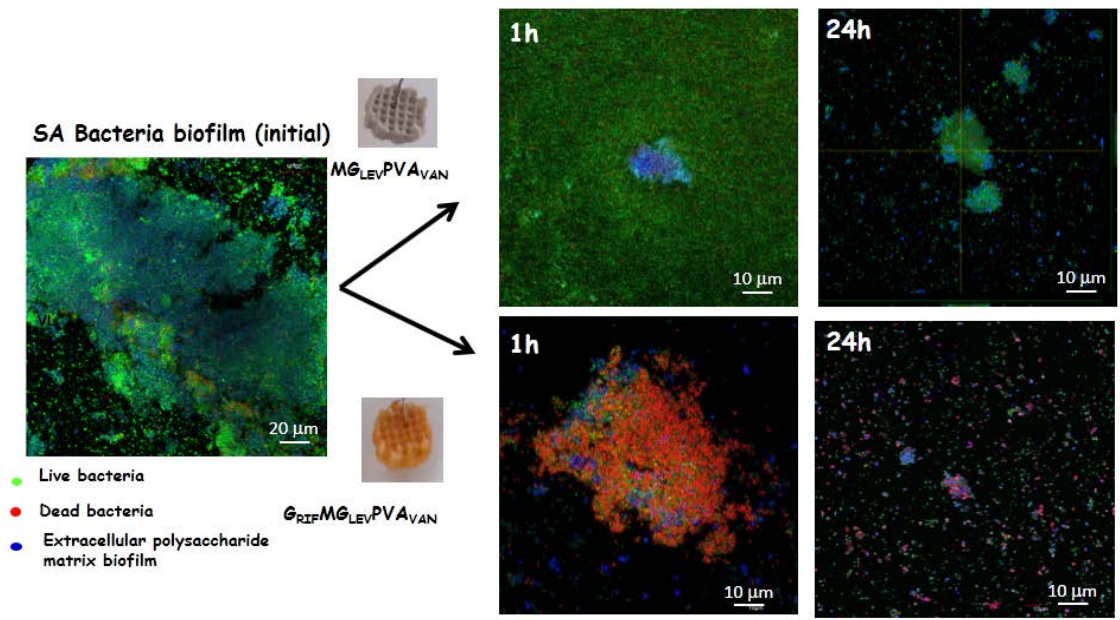

**Fig.8** Antimicrobial activity of both 3D MG$_{LEV}$PVA$_{VAN}$ (containing LEV and VAN) and G$_{RIF}$MG$_{LEV}$PVA$_{VAN}$ (containing RIF, LEV and VAN) scaffolds onto preformed *S. aureus* biofilm. The confocal images show clear differences between both 3D scaffolds with and without RIF. For 3D MG$_{LEV}$PVA$_{VAN}$ scaffolds, still appear small fragments of biofilm of size of 20-30 μm, which are formed by lived bacteria (green) covered by matrix biofilm layer (blue) after 24 h. For 3D G$_{RIF}$MG$_{LEV}$PVA$_{VAN}$ scaffolds, it is evident their rapid antimicrobial effect, showing the total destruction of biofilm, appearing all dead bacteria (red) after 1 h and a few scattering formed by small fragments of protective biolayer (blue), dead bacteria (red) and small amount of live bacteria (green) after 24 h.

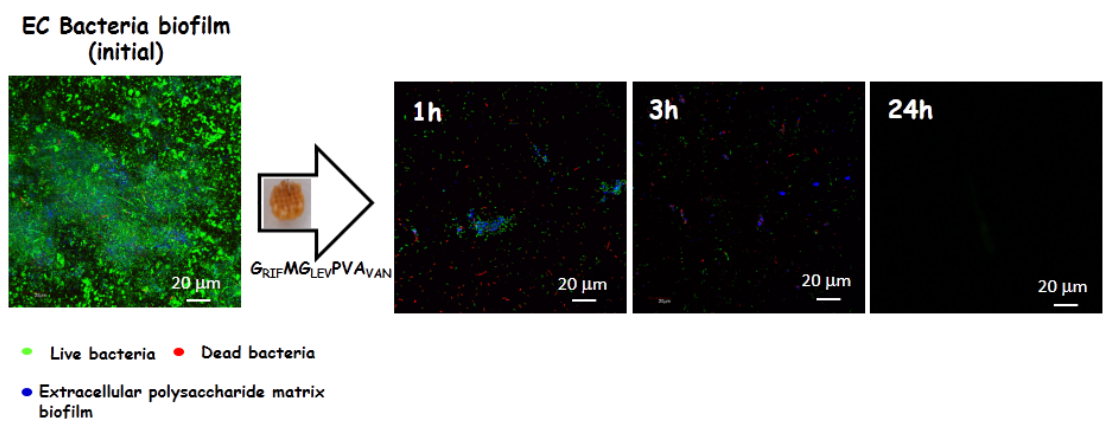

**Fig.9** Confocal microscopy study concerning to antimicrobial activity of G_RIF MG_LEV PVA_VAN (containing RIF, LEV and VAN) scaffolds onto Gram-negative E. coli (EC) biofilm. The confocal images show the initial biofilm preformed on covered glass-disk and after 1, 6 and 24 h of incubation with the 3D multidrug G_RIF MG_LEV PVA_VAN sample.

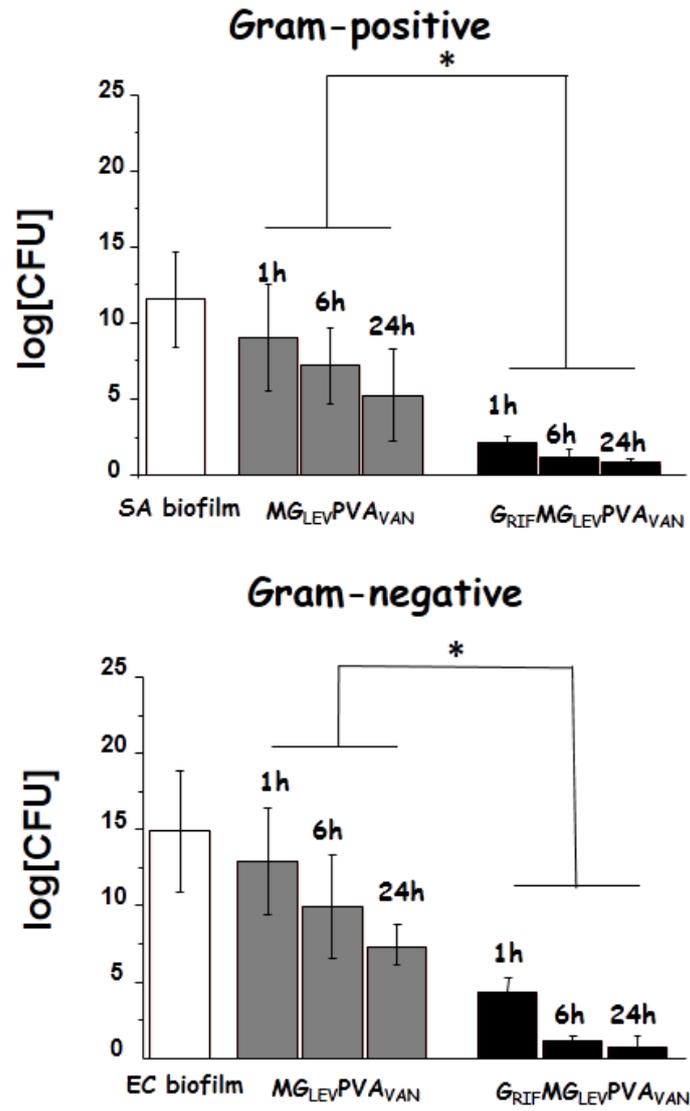

**Fig.10** Histograms showing the Log[CFU] from Gram-positive (*S. aureus*) and Gram-negative (E. coli) biofilms, respectively before and after incubation different times with both 3D MG$_{LEV}$PVA$_{VAN}$ and G$_{RIF}$MG$_{LEV}$PVA$_{VAN}$ scaffolds.

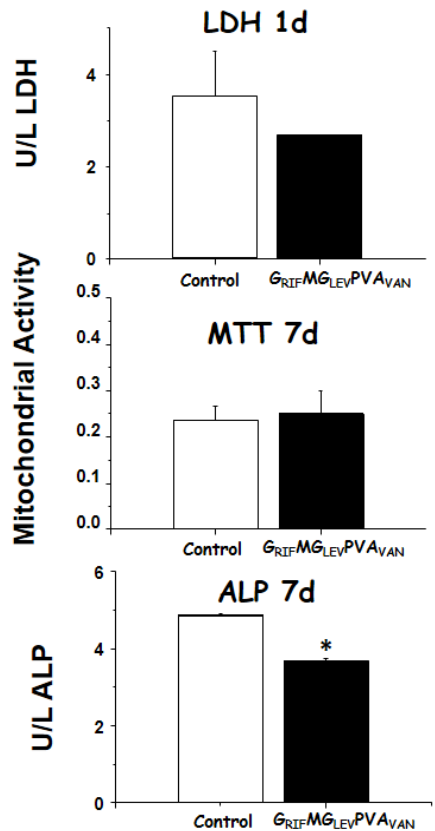

**Fig.11** Preliminary *in vitro* biocompatibility assays: (A) Cytotoxicity by lactate dehydrogenase (LDH) released into the medium after 1 day of incubation; (B) Proliferation assays in terms of mitochondrial activity (MTT) after 7 days incubation and (C) Differentiation assays in term of alkaline phosphatase activity (ALP) after 7 days of incubation of MC3T3-E1 preosteoblastic cells cultured onto G$_{RIF}$MG$_{LEV}$PVA$_{VAN}$ scaffolds. The values shown are means ± standard errors of a representative of three similar experiments carried out in duplicate. Differences between substrates at a given time point are not significant (p < 0.05, two-way ANOVA multiple comparison unless denoted by an asterisk (*).

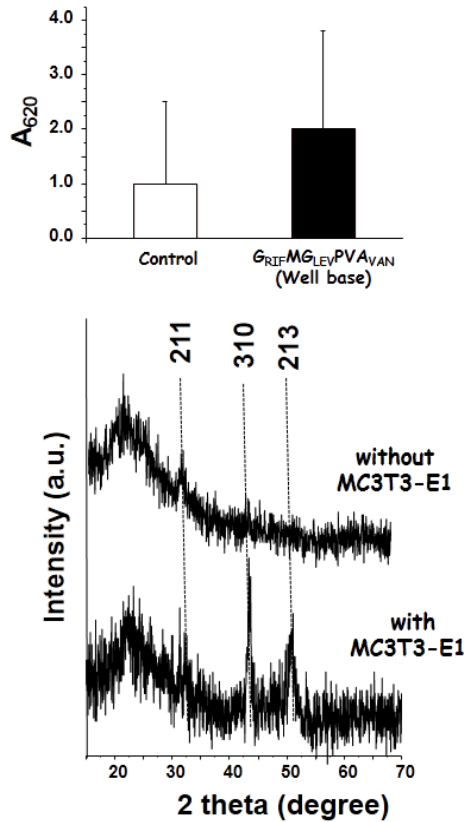

**Fig.12** Mineralization *in vitro* assays of MC3T3-E1 preosteoblastic cells cultured after 10 days of incubation with 3D G$_{RIF}$MG$_{LEV}$PVA$_{VAN}$ scaffolds. (Top) Alizarin assays measured in the well base showing no significant differences with the control. (Bottom) Surface characterization with XRD with and also without presence of preosteoblast cells XRD of 3D G$_{RIF}$MG$_{LEV}$PVA$_{VAN}$ scaffolds. Differences between substrates at a given time point are not significant ($p < 0.05$, two-way ANOVA multiple comparison unless denoted by an asterisk (*).

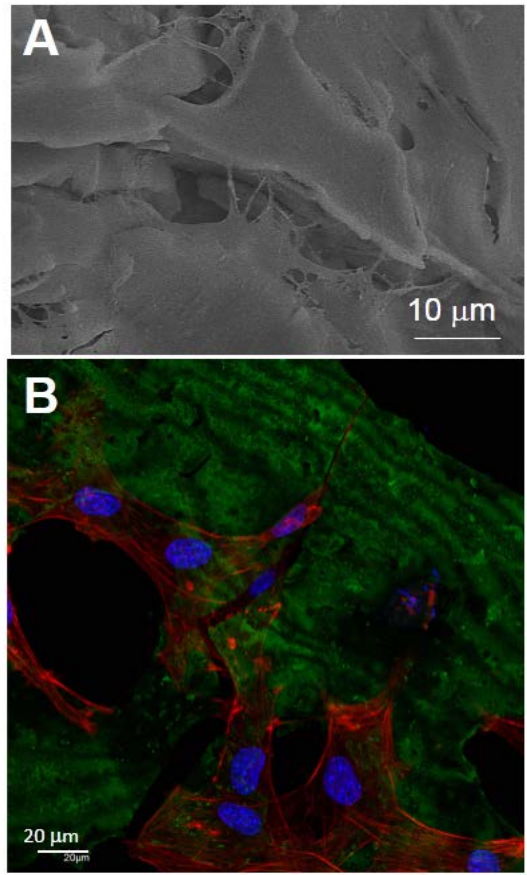

**Fig.13** (A) SEM micrograph and (B) Confocal image, showing the actin stained with Atto 565-conjugated phalloidin (red) and nuclei stained with DAPI (blue) of MC3T3-E1 preosteoblastic cells cultured on GEL$_{RIF}$MGHA$_{LEVO}$PVA$_{VAN}$ scaffold (reflected in green) after 7 days of incubation.

**Supporting information:**

**Table S1:** Elemental analyses data showing the percent of carbon, hydrogen and nitrogen of each table

| Sample | %C | %H | %N |
|---|---|---|---|
| **VAN*** | 43.85 | 5.64 | 7.06 |
| **LEV*** | 58.36 | 5.56 | 11.31 |
| **MG Powder** | 0.19 | 0.29 | 0.06 |
| **MGPVA** | 10.66 | 2.38 | 0.09 |
| **MG$_{LEV}$PVA** | 10.37 | 2.16 | 0.34 |
| **MGPVA$_{VAN}$** | 12.18 | 2.53 | 0.13 |
| **MG$_{LEV}$PVA$_{VAN}$** | 10.27 | 2.07 | 0.47 |

*Commercial drug powder from Sigma-Aldrich

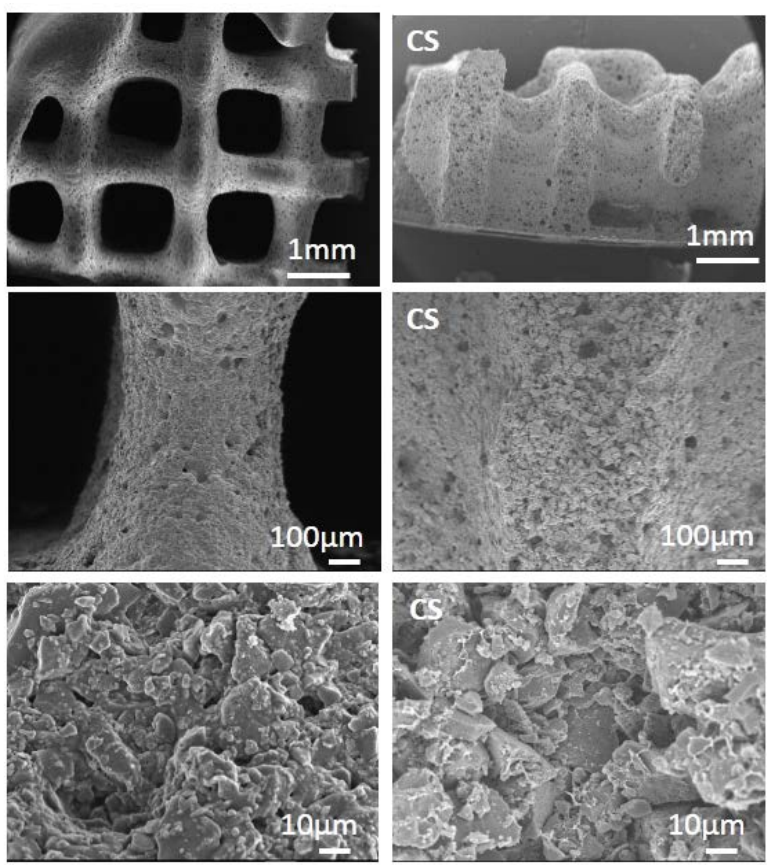

**Fig.S1** SEM study corresponding to MG$_{LEV}$PVA$_{VAN}$ scaffolds. Left shows different magnifications of the surface. Right shows the cross section view at different magnification.

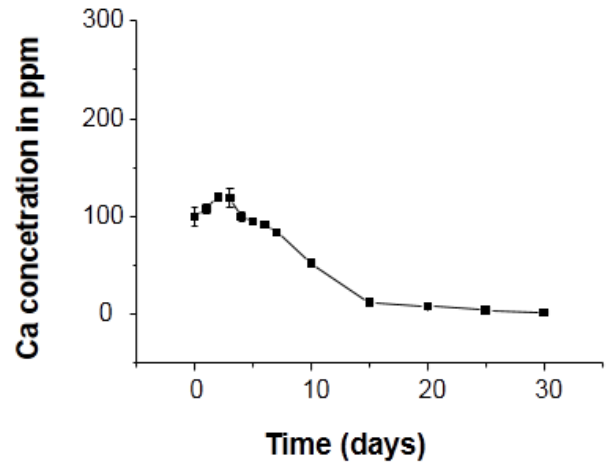

**Fig.S2** Variation of calcium concentration versus time when the G<sub>RIF</sub>MG<sub>LEV</sub>PVA<sub>VAN</sub> scaffolds in SBF showing a short increase of calcium followed to a notable decreasing. Note that the SBF was removed each 3 days.

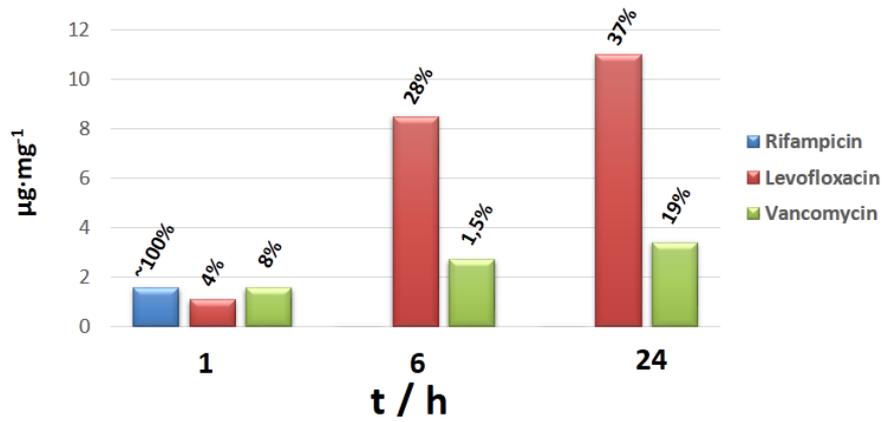

**Fig.S3** Dosage of each antimicrobial agent after 1, 6 and 24 h.

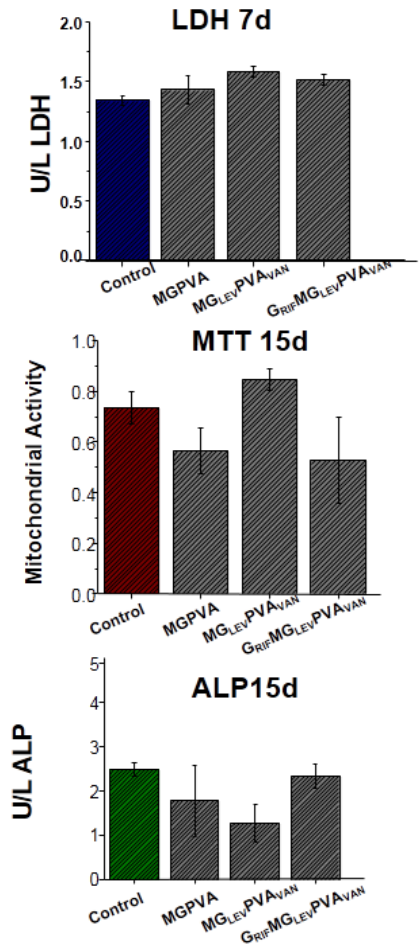

**Fig.S4** Long time *in vitro* of biocompatibility assays by incubation of G_RIF MG_LEV PVA_VAN scaffolds on MC3T3-E1 preosteoblastic cells cultured. Lactate dehydrogenase (LDH) determination after 7 days of incubation. Mitochondrial activity (MTT) after 15 days incubation and alkaline phosphatase activity (ALP) after 15 days of incubation. Not significant differences are shown.